\documentclass[conference]{IEEEtran}
\IEEEoverridecommandlockouts

\usepackage{cite}
\usepackage{amsmath,amssymb,amsfonts}
\usepackage{algorithmic}
\usepackage{graphicx}
\usepackage{textcomp}
\usepackage{xcolor}
\usepackage{tikz}
\usetikzlibrary{arrows.meta,positioning,fit,calc}
\usepackage{subcaption}
\usepackage{standalone}
\usepackage{tcolorbox}
\usepackage{paralist}

\usepackage{enumitem}
\usepackage{hyperref}
\usepackage{url}
\usepackage{multirow}
\usepackage{fancyvrb}
\newcounter{myenumi}
\renewcommand{\themyenumi}{(\roman{myenumi})}

\def\BibTeX{{\rm B\kern-.05em{\sc i\kern-.025em b}\kern-.08em
    T\kern-.1667em\lower.7ex\hbox{E}\kern-.125emX}}
\begin{document}

\typeout{Prism: Using main.tex as the root file. If your project is compiling a different file (e.g., IEEE-conference-template-062824.tex), set main.tex as the root/main file in Prism.}

\title{SpecLoop: An Agentic RTL-to-Specification Framework with Formal Verification Feedback Loop}
\author{Fu-Chieh Chang$^{1}$, Yu-Hsin Yang$^{1}$, Hung-Ming Huang$^{1}$,   Yun-Chia Hsu$^{2}$, Yin-Yu Lin$^{2}$, \\ Ming-Fang Tsai$^{2}$, Chun-Chih Yang$^{2}$, Pei-Yuan Wu$^{1}$ 
\\
{\small $^1$Graduate Institute of Communication Engineering, National Taiwan University, Taipei, Taiwan} \\
{\small $^2$MediaTek Inc, Hsinchu, Taiwan }
}

\maketitle

\begin{abstract}
RTL implementations frequently lack up-to-date or consistent specifications, making comprehension, maintenance, and verification costly and error-prone. While prior work has explored generating specifications from RTL using large language models (LLMs), ensuring that the generated documents faithfully capture design intent remains a major challenge. We present SpecLoop, an agentic framework for RTL-to-specification generation with a formal-verification-driven iterative feedback loop. SpecLoop first generates candidate specifications and then reconstructs RTL from these specifications; it uses formal equivalence checking tools between the reconstructed RTL and the original design to validate functional consistency. When mismatches are detected, counterexamples are fed back to iteratively refine the specifications until equivalence is proven or no further progress can be made. Experiments across multiple LLMs and RTL benchmarks show that incorporating formal verification feedback substantially improves specification correctness and robustness over LLM-only baselines, demonstrating the effectiveness of verification-guided specification generation.
\end{abstract}


\section{Introduction}
Large language models (LLMs) have recently shown strong capabilities in hardware design automation, covering RTL generation from specifications as well as RTL debugging and rewriting.
Prior works~\cite{tsai2024rtlfixer,pinckney2025rtlrewrite,ho2025verilogcoder,yu2025spec2rtl,10.1109/DAC63849.2025.11133191} improve RTL generation quality through an agentic, tool-augmented framework, providing semantics-level signals for evaluating and correcting generated designs. In the reverse direction, RTL-to-text tasks such as RTL understanding have also been studied~\cite{pinckney2025comprehensive,liu2025deeprtl,liu2025deeprtl2}, and early efforts have begun to explore RTL-to-specification generation~\cite{he2026refining,li2025specllm,huang2025assessing}.

\begin{figure}[!t]
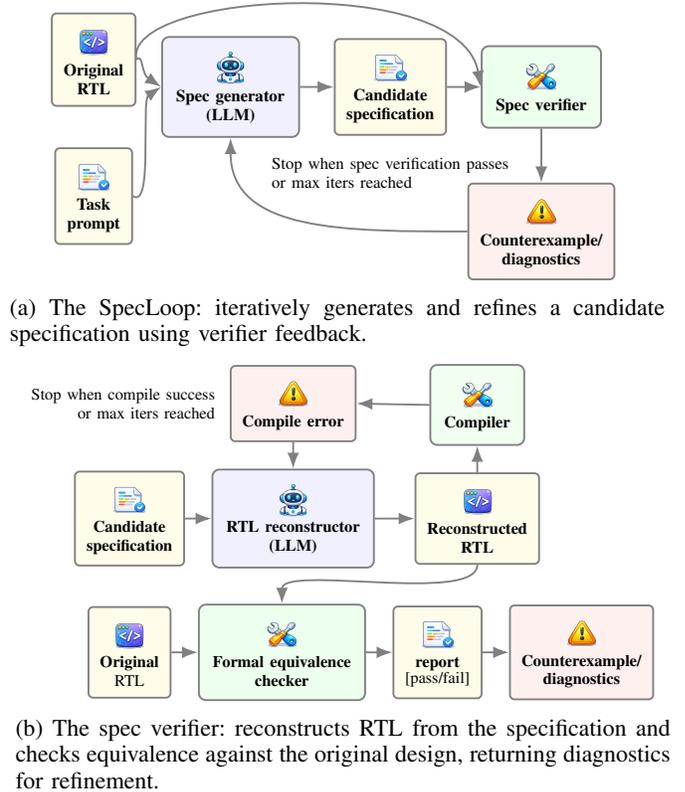

\centering
\begin{subfigure}[t]{0.48\textwidth}
  \centering
  \resizebox{0.9\linewidth}{!}{
    \input{figures/figure_architecture_specwriter.tex}
   }
  \caption{The SpecLoop: iteratively generates and refines a candidate specification using verifier feedback.}\label{fig:spec_loop}
\end{subfigure}\hfill
\vspace{0.2cm}
\begin{subfigure}[t]{0.48\textwidth}
  \centering
 \resizebox{\linewidth}{!}{
  \input{figures/figure_architecture_verifier.tex}
  }
  \caption{The spec verifier: reconstructs RTL from the specification and checks equivalence against the original design, returning diagnostics for refinement.}\label{fig:spec_verifier}
\end{subfigure}
\caption{The architecture of SpecLoop and the spec verifier.}
\label{fig:system_arch}
\end{figure}

However, generating high-quality specifications from RTL remains challenging: specifications require high-level abstraction of design intent and constraints, yet their quality is hard to measure and improve automatically. Inspired by the reconstruction-based evaluation protocol of Huang et al.~\cite{huang2025assessing}, 
we propose an \emph{agentic framework that integrates LLMs with formal equivalent checking tools}, enabling verification-guided RTL-to-specification generation that iteratively refines a candidate specification using feedback from RTL reconstruction and formal equivalence checking against the original design. Figure~\ref{fig:system_arch} provides an overview of the architecture.
Our core contributions are:
\begin{itemize}[leftmargin=*]
  \item We introduce an agentic RTL-to-specification generation framework driven by a verification-guided loop, leveraging formal equivalence checking as semantics-level feedback.
  \item We develop a strategy that separates compilation errors and functional mismatches, then uses actionable diagnostics to improve the specification.
  \item Our experiments show the framework benefits multiple LLMs and RTL benchmarks, outperforming prior work~\cite{huang2025assessing}. The approach achieves \emph{state-of-the-art} RTL-to-specification performance.
\end{itemize}
Overall, our agentic verification-guided approach turns specification generation into an \emph{iterative, checkable} process, enabling systematic improvement of spec quality beyond single-round prompting. 

\section{Related Work}
\subsection{Agentic Tool-Use and Iterative Refinement in LLMs}
Recent work shows that LLM inference improves with external tools and feedback. Agentic tool-augmented methods invoke compilers and analyzers~\cite{wu-etal-2024-toolplanner,zhang2024codeagent,wu2024autogen}, which are particularly effective in verifiable domains like programming.
Separately, LLMs can enhance responses via iterative refinement---reflection or self-correction~\cite{shah2025rethinking,lee2026intrinsic,chang2025unveiling}. Whether feedback is self- or tool-generated, both paradigms replace single-pass generation with multi-step error reduction.
These findings motivate our approach: tool-derived signals (compilation, equivalence checking) offer objective feedback, and iterative refinement improves results across rounds.

\subsection{Agentic Tool Use for LLM-Based RTL Generation}
Recent work has advanced agentic LLM frameworks that leverage external tools for RTL generation, debugging, and rewriting. Compilation and simulation tools are used to improve RTL generation~\cite{ho2025verilogcoder,yu2025spec2rtl,10.1109/DAC63849.2025.11133191} and syntax repair~\cite{tsai2024rtlfixer}. Equivalence checking tools are also integrated to rewrite RTL while preserving behavior for better PPA~\cite{pinckney2025rtlrewrite}, and to support dataset construction~\cite{yubeaton2025verithoughts} and evaluation~\cite{jin2025realbenchbenchmarkingveriloggeneration} in spec-to-RTL workflows. However, these methods do not transfer straightforwardly to the reverse task---RTL-to-specification---which remains challenging because specification quality is difficult to evaluate, limiting their direct applicability.

\subsection{RTL Understanding and Specification Generation}
Prior RTL-to-text studies largely split into RTL understanding and specification generation. RTL understanding produces narrative, often line-by-line explanations, whereas specification generation requires higher-level, structured semantics (e.g., explicit port names). Table~\ref{tab:understanding_vs_spec} gives a concrete contrast.
\begin{table}[!t]
\centering
\begin{scriptsize}
\caption{Illustrative contrast between RTL understanding and specification generation outputs of a given RTL snippet.}
\label{tab:understanding_vs_spec}
\begin{tabular}{|p{0.18\linewidth}| p{0.72\linewidth}|}
\hline
\textbf{Type} & \textbf{Example} \\
\hline
\textbf{RTL snippet} & \texttt{module HalfAdder (input a, input b, output sum, output cout); assign \{cout, sum\} = a+b; endmodule} \\
\hline
\textbf{RTL understanding (narrative)} & This Verilog code defines a module named \texttt{HalfAdder} that implements a basic half adder circuit. The module takes two single-bit inputs (\texttt{a} and \texttt{b}) and produces two outputs: \texttt{sum} (the result of the addition) and \texttt{cout} (carry out). The assign statement uses concatenation to calculate both the sum bit and carry bit from the binary addition of inputs a and b. \\
\hline
\textbf{Specification generation (structured)} & \textbf{Top Module:} \texttt{HalfAdder}. \textbf{Inputs:} \texttt{a} (1-bit), \texttt{b} (1-bit). \textbf{Outputs:} \texttt{sum} (1-bit), \texttt{cout} (1-bit). \textbf{Functional Description:} This module implements a half adder circuit that performs binary addition of two 1-bit operands. The module takes two single-bit inputs and produces a 1-bit sum output along with a 1-bit carry-out output. \\
\hline
\end{tabular}
\end{scriptsize}
\end{table}
We summarize related work below.
\begin{inparaenum}[(i)]
\item \textbf{RTL understanding.} Pinckney et al.~\cite{pinckney2025comprehensive} introduced \textit{CVDP}. Liu et al.~\cite{liu2025deeprtl,liu2025deeprtl2} proposed \textit{DeepRTL} and \textit{DeepRTL2}. Both mainly target code-level comprehension, not spec derivation; Pinckney et al.~\cite{pinckney2025comprehensive} further note spec generation is a more challenging problem than comprehension.
\item \textbf{Specification generation.} He et al.~\cite{he2026refining} use synthetic specs to improve RTL generation, but they focus on RTL quality. SpecLLM~\cite{li2025specllm} categorizes spec modalities but remains largely qualitative. Huang et al.~\cite{huang2025assessing} evaluate RTL-to-spec generation extensively, yet focus only on benchmarking.
\end{inparaenum}
Unlike prior work, we are the first to explicitly study how to improve spec quality using formal-verification-guided iterative refinement.

\section{System Architecture}
Inspired by Huang et al.~\cite{huang2025assessing}, we use an iterative reconstruction-and-check method. 
Figure~\ref{fig:system_arch} summarizes our agentic verification-guided RTL-to-specification framework. 
\begin{inparaenum}
\item[] Fig.~\ref{fig:spec_loop} shows the architecture of SpecLoop, where the \textbf{spec generator} produces a \textbf{candidate specification} from the \textbf{input RTL code} and \textbf{task prompt}. This candidate specification is then passed to a \textbf{spec verifier} to check its correctness. If verification fails, the resulting \textbf{counterexamples and diagnostics} are fed back to the spec generator to produce a revised specification. This process iterates until equivalence is proven or the maximum iteration budget is reached.
\item[] Fig.~\ref{fig:spec_verifier} shows the architecture of the \textbf{spec verifier}. It translates the \textbf{candidate specification} into \textbf{reconstructed RTL} via an \textbf{RTL reconstructor}, then compiles it. If compilation fails, the \textbf{compile errors} are fed back to the reconstructor to regenerate a compilable design. The compilable RTL is then compared with the original RTL using a \textbf{formal equivalence checker}, which returns a \textbf{report} (pass/fail). On failure, the \textbf{counterexamples and diagnostics} are fed back to the spec generator to refine the specification.
\end{inparaenum}

\begin{figure*}[!t]
\centering
\begin{scriptsize}
  \begin{tabular}[t]{c c}
\resizebox{.48\textwidth}{!}{
\input{figures/figure_prompt_template_1.tex}
}
 & 
 \resizebox{.48\textwidth}{!}{
\input{figures/figure_prompt_template_2.tex}
 }
\end{tabular}
\caption{Multi-step prompt templates for the specification generator. \textbf{(Left)} First round: analyze RTL, write a structured specification, and self-check. \textbf{(Right)} Refinement: use verifier diagnostics (e.g., compiler errors or counterexamples) to edit only affected fields and keep the rest unchanged. Some line breaks are omitted due to page limits.}
\label{fig:prompt_template}
\end{scriptsize}
\end{figure*}

\subsection{Spec Generator}
Huang et al.~\cite{huang2025assessing} show that prompt design strongly influences RTL-to-spec generation. We adapt and extend their multi-step reasoning prompt to enable reasoning and self-reflection before finalizing the specification and to enforce a structured output format, as illustrated in Fig.~\ref{fig:prompt_template}. 
\begin{inparaenum}[(1)]
\item \textbf{The left block} shows our first-round prompt. It uses three steps: (a) analyze the RTL (signals, data/control flow, clock/reset); (b) write a structured specification with fixed fields; and (c) self-check for completeness and RTL consistency. 
\item \textbf{The right block} shows our refinement-round prompt. It asks the model to (a) summarize the verification report, (b) identify the likely root cause (see Sec.~\ref{sec:rtl_reconstructor} and Sec.~\ref{sec:formal_verification_tool}), and (c) decide whether (and how) to revise only the affected specification fields. This makes feedback actionable and enables incremental improvements.
\end{inparaenum}
At the bottom of these prompts, the output rules make our reconstruction-and-check pipeline more robust. First, explicit reasoning steps improve correctness and interpretability. Second, markers like \texttt{[SPEC\_START]} and \texttt{[SPEC\_END]} keep specs easy to parse and tolerant of extra text. Third, forbidding Verilog code in specs pushes the model toward generating abstract and concise specs.

\subsection{RTL Reconstructor}\label{sec:rtl_reconstructor}
We use an LLM-based RTL reconstructor to turn each natural-language specification into compilable RTL. Although this step enables semantics-level validation, reconstruction may fail: the model can misinterpret underspecified requirements, miss corner cases, or produce syntactically invalid RTL. As a result, an equivalence-check failure cannot be attributed solely to the spec generator.
To mitigate this confounder, we separate \emph{reconstruction validity} from \emph{functional equivalence}. Before formal equivalence checking (FEC), we compile the reconstructed RTL with a lightweight compiler; if compilation fails, we rerun reconstruction based on compile errors. If it compiles yet equivalence still fails, we consider the specification incomplete or incorrect and feed back counterexamples for refinement. Even so, reconstructor capacity may still bound end-to-end performance, including cases where the specification is correct (see Sec.~\ref{sec:formal_verification_tool} and Sec.~\ref{sec:limitations}).

\subsection{Formal Equivalence Checker}\label{sec:formal_verification_tool}
We use a FEC tool to compare the original RTL with the reconstructed RTL. Compared to simulation, equivalence checking provides an exhaustive, semantics-level criterion: when it succeeds, it shows the specification is detailed enough for the reconstructed RTL to be functionally equivalent to the original. When the check fails, it may do so for several reasons:
\begin{enumerate}[label=\textbf{E.\arabic*},ref=E.\arabic*, leftmargin=*]
  \item \textbf{Invalid original RTL.} If the original RTL fails to compile or cannot be parsed by the FEC tool, the checker may be unable to produce a meaningful verdict. In this case, we stop the iteration.
  \label{err:E1}
  \item \textbf{Non-compilable reconstructed RTL.} If the reconstructed RTL does not compile, we regard it as a reconstruction failure and re-run the reconstructor with compiler errors. Since such errors can also stem from spec inaccuracy (e.g., missing widths), after a fixed retry budget we attribute persistent failures to specification quality and feed the compile error to the specification generator.
  \label{err:E2}
  \item \textbf{Functional mismatch.} A true mismatch yields a counterexample where output signals diverge. Since mismatches can also stem from reconstructor imperfections, we may repeat reconstruction up to a fixed budget; if mismatches persist, we feed back the counterexample to the specification generator to amend the specification.
  \label{err:E3}
  \item \textbf{Inconclusive failures.} FEC may return inconclusive results due to timeouts or other tool issues. We first re-run the reconstructor; if inconclusive outcomes persist, we attribute them to an ill-posed specification and send an error message to the spec generator requesting refinement.
  \label{err:E4}
\end{enumerate}
\textbf{Design choice---information hiding:} To fairly assess the information contained in the specification, we deliberately do \emph{not} feed FEC logs back to the reconstructor, because they can reveal details of the original RTL (an input to the checker) and effectively bypass the specification. The only exception is compilation failures: we provide compiler errors to the reconstructor because they do not directly leak original RTL semantics and are necessary to produce a compilable design. We summarize the actions to the reconstructor and spec generator in Table~\ref{tab:verifier_actions}.

\begin{table}[!t]
\centering
\begin{scriptsize}
\caption{Verifier error handling policy.}
\label{tab:verifier_actions}
\begin{tabular}{|p{0.05\linewidth}| p{0.215\linewidth}| p{0.55\linewidth}|}
\hline
\textbf{Error} & \textbf{To Reconstructor} & \textbf{To Spec Generator} \\
\hline\hline
E.1 & Stop iteration & Non-verifiable; no feedback loop \\ \hline
E.2 & Reconstruct with compile errors & After reconstruct budget, send compile error \\ \hline
E.3 & Reconstruct & After reconstruct budget, send counterexample \\ \hline
E.4 & Reconstruct & After reconstruct budget, send error message  \\ \hline
\end{tabular}
\end{scriptsize}
\end{table}

\section{Experiment Settings}\label{sec:exp}
We conduct experiments to evaluate our system along three questions:
\begin{inparaenum}[(1)]
\item whether verifier-guided refinement improves RTL-to-specification generation over a single-round baseline;
\item whether diagnostic feedback (compiler errors and counterexamples) is more effective than a binary pass/fail signal; and
\item how robust the pipeline is across different LLM backbones and benchmarks.
\end{inparaenum}
\subsection{LLM Candidates}
We have two LLM roles in our pipeline: a \emph{specification generator} and an \emph{RTL reconstructor}. In principle, using a stronger reconstructor could reduce reconstruction-induced errors and improve end-to-end success rates; however, it would also confound comparisons by injecting additional model capacity. To isolate the contribution of our verification-guided loop, we use \emph{exactly the same} LLM for both roles. Thus, any gain over single-round reflects verifier integration. Besides, our method is model-agnostic and requires no fine-tuning; we directly plug in existing LLMs. Accordingly, we evaluate a wide range of existing LLMs including Llama4-Scout, Llama4-Maverick, Qwen3-Coder-30B, Qwen3-Coder-480B, DeepSeek-v3.1, and GLM-4.6, and observe consistent gains across model sizes and architectures.

\subsection{Benchmarks and Metrics}
We evaluate on two representative RTL benchmarks: VerilogEval~\cite{liu2023verilogeval} and RTLLM~\cite{lu2024rtllm}.
VerilogEval contains toy-level modules, whereas RTLLM includes larger designs.
Since these benchmarks were originally designed for spec-to-RTL evaluation, we follow the RTL-to-spec protocol of Huang et al.~\cite{huang2025assessing}: for each input RTL, we generate a natural-language specification and evaluate it using the RTL Reconstruction Score (RR score)~\cite{huang2025assessing}. RR checks whether RTL reconstructed from the specification passes the testbenches, with reconstruction performed by GPT-5 Codex~\cite{openai_gpt5}. We choose RR over semantic-level alternatives such as GPT-Score~\cite{liu2025deeprtl} because it is more sensitive to specification flaws~\cite{huang2025assessing} than semantic-level metrics. Although RR also includes an RTL-reconstruction step, it does not inherently favor our method:
\begin{inparaenum}[(1)]
\item RR uses frontier reconstructors (e.g., the GPT-5 series) for reconstruction fidelity, whereas SpecLoop uses the same (often weaker) model for both generation and reconstruction to avoid confounding extra capacity; and
\item RR evaluates reconstructed RTL with testbenches rather than formal equivalence, measuring behavior under the provided environments rather than our equivalence-checking objective.
\end{inparaenum}

\subsection{Baselines}
We consider two baselines. 
\begin{inparaenum}[(1)]
  \item 
\textbf{Single Round:} This follows prior work~\cite{huang2025assessing}: the LLM produces a specification in a single pass from the input RTL, without verifier-guided refinement. We reproduce their results with a minor prompt change (enabling intermediate reasoning outputs) to make the outputs more interpretable.
 \item \textbf{Pass/Fail-only verifier:} we run the same reconstruction and equivalence-checking pipeline, but the verifier only returns a binary decision (pass/fail) without diagnostic messages (e.g., compiler errors or counterexamples). On failure, we prompt the LLM to retry with no extra signals, using the same retry budget as the full-feedback method.
\end{inparaenum}

\subsection{Other Implementation Details}
For equivalence checking, we use Yosys EQY~\cite{yosys_eqy} with a SAT backend and a depth bound of 10. We build a lightweight log parser that maps EQY outcomes to error types (\ref{err:E1}--\ref{err:E4}), and a VCD parser that extracts counterexamples for \ref{err:E3}. When error \ref{err:E1} occurs, we terminate the feedback loop since the case cannot benefit from verifier feedback. We allow up to two retries for the spec generator, for three spec generations total including the initial round. Each spec allows up to two RTL reconstructor retries. We call LLM APIs through OpenRouter~\cite{openrouter} and use a temperature of 0.4 to trade off stochastic decoding and correctness.
To reduce statistical variance, we run each experiment three times and report the mean and standard deviation.

\begin{table}[!t]
\centering
\begin{scriptsize}
\caption{Experimental results (mean $\pm$ std) on VerilogEval and RTLLM with RR-Score; darker tone indicates higher ranking. Verifier-guided variants (Pass/Fail-Only and Full Diagnosis) outperform the Single-Round baseline; overall, Full Diagnosis ranks best, with Pass/Fail-Only ranks second.}
\label{tab:exp_result}
\resizebox{.49\textwidth}{!}{
\begin{tabular}{|c|ccc|}
\hline
\multicolumn{1}{|c|}{Model\textbackslash Method}      & \begin{tabular}[c]{@{}c@{}}Single Round\\ (\cite{huang2025assessing})\end{tabular} & \begin{tabular}[c]{@{}c@{}}Pass/Fail-Only\\ (Ours)\end{tabular} & \begin{tabular}[c]{@{}c@{}}Full Diagnosis\\ (Ours)\end{tabular} \\ 
\hline\hline
\multicolumn{4}{|c|}{Dataset: VerilogEval}      \\
\hline

Qwen3-Coder-30B & $\textcolor{black!30}{\mathbf{0.722 \pm 0.016}}$ & $\textcolor{black!60}{\mathbf{0.759 \pm 0.017}}$ & $\textcolor{black}{\mathbf{0.795 \pm 0.021}}$ \\
Qwen3-Coder-480B & $\textcolor{black!60}{\mathbf{0.885 \pm 0.016}}$ & $\textcolor{black!30}{\mathbf{0.878 \pm 0.009}}$ & $\textcolor{black}{\mathbf{0.912 \pm 0.012}}$ \\
DeepSeek-V3.1 & $\textcolor{black!30}{\mathbf{0.865 \pm 0.019}}$ & $\textcolor{black!60}{\mathbf{0.908 \pm 0.011}}$ & $\textcolor{black}{\mathbf{0.940 \pm 0.003}}$ \\
Llama4-Scout-16E & $\textcolor{black!30}{\mathbf{0.686 \pm 0.009}}$ & $\textcolor{black}{\mathbf{0.722 \pm 0.016}}$ & $\textcolor{black!60}{\mathbf{0.705 \pm 0.019}}$ \\
Llama4-Maverick-128E & $\textcolor{black!60}{\mathbf{0.818 \pm 0.006}}$ & $\textcolor{black!30}{\mathbf{0.812 \pm 0.017}}$ & $\textcolor{black}{\mathbf{0.842 \pm 0.015}}$ \\
GLM-4.6 & $\textcolor{black!30}{\mathbf{0.906 \pm 0.021}}$ & $\textcolor{black!60}{\mathbf{0.921 \pm 0.006}}$ & $\textcolor{black}{\mathbf{0.934 \pm 0.020}}$ \\
\hline\hline
\multicolumn{4}{|c|}{Dataset: RTLLM}      \\
\hline
Qwen3-Coder-30B & $\textcolor{black!30}{\mathbf{0.733 \pm 0.009}}$ & $\textcolor{black}{\mathbf{0.747 \pm 0.019}}$ & $\textcolor{black}{\mathbf{0.747 \pm 0.019}}$ \\
Qwen3-Coder-480B & $\textcolor{black!60}{\mathbf{0.767 \pm 0.025}}$ & $\textcolor{black}{\mathbf{0.787 \pm 0.019}}$ & $\textcolor{black!60}{\mathbf{0.767 \pm 0.041}}$ \\
DeepSeek-V3.1 & $\textcolor{black!30}{\mathbf{0.767 \pm 0.025}}$ & $\textcolor{black!60}{\mathbf{0.793 \pm 0.025}}$ & $\textcolor{black}{\mathbf{0.800 \pm 0.016}}$ \\
Llama4-Scout-16E & $\textcolor{black!30}{\mathbf{0.667 \pm 0.025}}$ & $\textcolor{black}{\mathbf{0.727 \pm 0.034}}$ & $\textcolor{black!60}{\mathbf{0.687 \pm 0.025}}$ \\
Llama4-Maverick-128E & $\textcolor{black!60}{\mathbf{0.793 \pm 0.034}}$ & $\textcolor{black!30}{\mathbf{0.767 \pm 0.009}}$ & $\textcolor{black}{\mathbf{0.813 \pm 0.009}}$ \\
GLM-4.6 & $\textcolor{black!30}{\mathbf{0.813 \pm 0.034}}$ & $\textcolor{black!60}{\mathbf{0.847 \pm 0.009}}$ & $\textcolor{black}{\mathbf{0.880 \pm 0.000}}$ \\
\hline
\end{tabular}
}
\end{scriptsize}
\end{table}

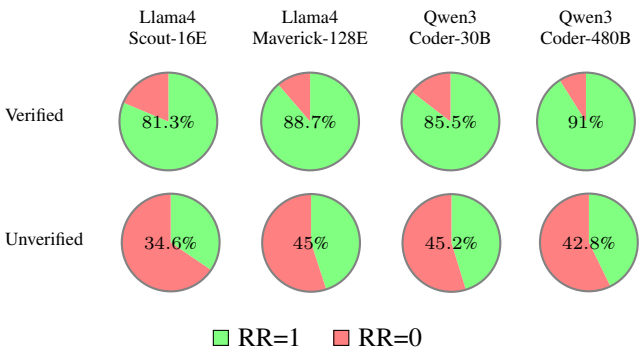
\begin{figure}[t]
  \centering
  \definecolor{vRR1}{RGB}{128,255,128}
\definecolor{vRR0}{RGB}{255,128,128}
\definecolor{vrBorder}{RGB}{130,130,130}

\newcommand{\PieCell}[2]{%
  \begin{tikzpicture}[baseline={(current bounding box.center)}]
    \def\r{0.65cm}
    \pgfmathsetmacro{\angV}{360*#1}
    \pgfmathsetmacro{\pct}{100*#1}
    \path[fill=vRR1] (0,0) -- (90:\r) arc (90:{90-\angV}:\r) -- cycle;
    \path[fill=vRR0] (0,0) -- ({90-\angV}:\r) arc ({90-\angV}:{90-360}:\r) -- cycle;
    \draw[line width=0.8pt, draw=vrBorder] (0,0) circle (\r);
    \node[align=center] at (0,0) {\scriptsize\pgfmathprintnumber[fixed,precision=1]{\pct}\%};
  \end{tikzpicture}%
}
\begin{tikzpicture}[font=\scriptsize]
  \node at (0,0) {%
    \begin{tabular}{@{}lcccc@{}}
      &
      {\begin{tabular}[c]{@{}c@{}}Llama4\\ Scout-16E\end{tabular}} &
      {\begin{tabular}[c]{@{}c@{}}Llama4\\ Maverick-128E\end{tabular}} &
      {\begin{tabular}[c]{@{}c@{}}Qwen3\\ Coder-30B\end{tabular}} &
      {\begin{tabular}[c]{@{}c@{}}Qwen3\\ Coder-480B\end{tabular}}  
      \\
      \\ 
      Verified &
      \PieCell{0.8132118451025057}{0.1867881548974943} &
      \PieCell{0.8871635610766045}{0.11283643892339544} &      
      \PieCell{0.8546712802768166}{0.145328719723184} &
      \PieCell{0.91015625}{0.08984375} \\
      \\
      Unverified &
      \PieCell{0.34563758389261745}{0.6543624161073825} &
      \PieCell{0.44976076555023925}{0.5502392344497608} &      
      \PieCell{0.4523809523809524}{0.5476190476190477} &
      \PieCell{0.4276315789473684}{0.5723684210526315} 
      \\
    \end{tabular}%
  };
\end{tikzpicture}


\vspace{1mm}
\begin{tabular}{@{}ll@{}}
  \tikz{\fill[vRR1] (0,0) rectangle (2mm,2mm); \draw (0,0) rectangle (2mm,2mm);} RR=1  &
  \tikz{\fill[vRR0] (0,0) rectangle (2mm,2mm); \draw (0,0) rectangle (2mm,2mm);}  RR=0 \\
\end{tabular}
  \caption{Ratio of verified specifications vs. unverified for different models and RR scores averaged over benchmarks and verifier variants. Verified specs show higher RR=1 proportions than unverified ones across models, especially for stronger models.}
  \label{fig:verified_ratio}
\end{figure}

\begin{figure*}[!t]
  \centering
 \begin{scriptsize}
  \begin{tabular}[t]{c c}
  \resizebox{0.48\textwidth}{!}{
        \input{figures/qualitative_analysis_1.tex}
  }
  &
  \resizebox{0.48\textwidth}{!}{
    \input{figures/qualitative_analysis_2.tex}
  }
  \end{tabular}
  \end{scriptsize}
  \caption{Qualitative Analysis. Selected text segments are highlighted in red for clarity. This example shows how SpecLoop fixes a spec error: the first-round spec wrongly states an asynchronous reset, reconstruction then fails equivalence checking, and the diagnosis guides the next round to revise the spec to a synchronous reset, after which the verifier passes.}
  \label{fig:qualitative_analysis}
\end{figure*}

\section{Experiment Results}
\subsection{Overall Quantitative Results}
Table~\ref{tab:exp_result} reports the RTL reconstruction score (mean $\pm$ std) for three settings: \textit{Single Round}, \textit{Pass/Fail-Only verifier}, and \textit{Full Diagnosis}. We highlight the following observations.
\begin{inparaenum}[(i)]
\item \textbf{Iterative Loop consistently improves over Single-Round generation.} Across results, adding the verification loop (Pass/Fail-Only or Full Diagnosis) outperforms Single Round, indicating that verifier feedback---either a pass/fail outcome or full diagnostic information (ranking highest in 4 and 9 among the 12 model/dataset settings, respectively)---is a valuable inference-time signal.
\item \textbf{Even a Pass/Fail-only verifier can outperform Single-Round generation.} Despite lacking detailed diagnostics, Pass/Fail-only achieves more top rankings than Single-Round (4 vs.\ 0 across 12 model/dataset settings). This reflects observations from prior work~\cite{lee2026intrinsic,chang2025unveiling} that, without external guidance, repeatedly self-correcting and reflecting on reasoning steps or outcomes remains beneficial.
 \item \textbf{Richer diagnostics help the most.} Full Diagnosis delivers the strongest performance, indicating that detailed feedback (e.g., counterexamples) enables models to make more precise refinements. Overall, these results suggest that making RTL-to-spec generation \emph{iterative and checkable} improves end-to-end correctness and achieves \emph{state-of-the-art performance} over the prior single-round baseline.
 \end{inparaenum}

\subsection{Verified vs.\ Unverified Specifications}
We further analyze the impact of the spec verifier in the feedback loop. 
Figure~\ref{fig:verified_ratio} reports, for each model, the fraction of samples achieving $RR{=}1$ and $RR{=}0$ in two groups: specifications verified (i.e., passing the spec verifier's FEC) within the budget and those left unverified after budget exhaustion. Each pie chart summarizes the verified vs.\ unverified split, averaged over benchmarks and feedback variants (Pass/Fail-Only and Full Diagnosis).
Two observations follow. 
\begin{inparaenum}[(i)]
\item Verified specifications are more likely to achieve $RR{=}1$. 
\item Stronger RTL reconstructors yield a higher $RR{=}1$ ratio among verified specifications, as shown by Qwen3-Coder-480B and Llama4-Maverick compared with Qwen3-Coder-30B and Llama4-Scout. 
\end{inparaenum}
These observations suggest that verified specifications are higher quality (by RR score), with larger gains for stronger models.

\subsection{Qualitative Study}
Figure~\ref{fig:qualitative_analysis} presents a qualitative example. 
\begin{inparaenum}[(i)]
\item The original RTL design is a counter that increments from 0 to 999 with a synchronous reset. 
\item In the first LLM response, however, the model incorrectly infers that the reset is asynchronous. 
\item Our spec-verification process then reconstructs RTL from this incorrect specification; because the reconstructed design contains an asynchronous reset, the equivalence checker flags a mismatch. 
\item We feed the resulting diagnosis back to the LLM, and in the second round it correctly identifies that the mistake stems from mischaracterizing the reset behavior and revises the specification to use a synchronous reset. 
\item In the next verification round, the reconstructor produces a counter with a synchronous reset and the design passes equivalence checking. 
\end{inparaenum}
This example illustrates how the spec-verification loop can detect errors in generated specifications and guide the LLM to fix them.

\section{Limitations and Future Work}\label{sec:limitations}
Our work has several limitations that fall into two categories. \textbf{Fundamental limitations of the reconstruction-based paradigm and metric:} end-to-end results depend on the RTL reconstructor and the equivalence-checking setup, so failures in reconstruction or resource-bounded checking (e.g., fixed depth) can block useful feedback on harder designs. Moreover, evaluation with the RR score also depends on the reconstruction capabilities of frontier models.
\textbf{Engineering gaps that can be improved:} feedback parsing and prompting remain heuristic, and scaling to large industrial RTL codebases will require multi-agent system~\cite{ho2025verilogcoder,10.1109/DAC63849.2025.11133191,yu2025spec2rtl} (e.g., planning, tool selection, and handling IP dependencies). Future work includes strengthening reconstruction and verification pipelines and developing multi-agent system for handling more complex designs.

\section{Conclusion}
We presented a agentic verification-guided framework for RTL-to-specification generation that turns the problem into an iterative reconstruction-and-check loop. By reconstructing RTL from candidate specifications and using formal equivalence checking to produce semantics-level feedback, our approach systematically refines specifications beyond one-shot prompting. Experiments show verifier guidance improves spec quality, as evidenced by reconstruction success on frontier models, and that richer diagnostics are beneficial. We believe this direction can make RTL-to-spec generation a practical component in real hardware documentation workflows.

\bibliographystyle{ieeetr}
\bibliography{reference}

@inproceedings{liu2023verilogeval,
  title={Verilogeval: Evaluating large language models for verilog code generation},
  author={Liu, Mingjie and Pinckney, Nathaniel and Khailany, Brucek and Ren, Haoxing},
  booktitle={2023 IEEE/ACM International Conference on Computer Aided Design (ICCAD)},
  pages={1--8},
  year={2023},
  organization={IEEE}
}

@inproceedings{lu2024rtllm,
  title={Rtllm: An open-source benchmark for design rtl generation with large language model},
  author={Lu, Yao and Liu, Shang and Zhang, Qijun and Xie, Zhiyao},
  booktitle={2024 29th Asia and South Pacific Design Automation Conference (ASP-DAC)},
  pages={722--727},
  year={2024},
  organization={IEEE}
}

@inproceedings{ho2025verilogcoder,
  title={Verilogcoder: Autonomous verilog coding agents with graph-based planning and abstract syntax tree (ast)-based waveform tracing tool},
  author={Ho, Chia-Tung and Ren, Haoxing and Khailany, Brucek},
  booktitle={Proceedings of the AAAI Conference on Artificial Intelligence},
  volume={39},
  number={1},
  pages={300--307},
  year={2025}
}

@inproceedings{tsai2024rtlfixer,
  title={Rtlfixer: Automatically fixing rtl syntax errors with large language model},
  author={Tsai, YunDa and Liu, Mingjie and Ren, Haoxing},
  booktitle={Proceedings of the 61st ACM/IEEE Design Automation Conference},
  pages={1--6},
  year={2024}
}

@inproceedings{pinckney2025rtlrewrite,
  title={RTLRewriter: Methodologies for Large Models aided RTL Code Optimization},
  author={Yao, Xufeng and Wang, Yiwen and Li, Xing and Lian, Yingzhao and Chen, Ran and Chen, Lei and Yuan, Mingxuan and Xu, Hong and Yu, Bei},
  booktitle={2024 IEEE/ACM International Conference on Computer Aided Design (ICCAD)},
  year={2024}
}

@misc{yosys_eqy,
  title        = {Equivalence Checking with Yosys (EQY) Documentation},
  howpublished = {\url{https://yosyshq.readthedocs.io/projects/eqy/en/latest/}},
  note         = {Accessed: 2026-02-21}
}

@misc{openrouter,
  title        = {OpenRouter},
  howpublished = {\url{https://openrouter.ai/}},
  note         = {Accessed: 2026-02-21}
}

@article{pinckney2025comprehensive,
  title={Comprehensive Verilog Design Problems: A Next-Generation Benchmark Dataset for Evaluating Large Language Models and Agents on RTL Design and Verification},
  author={Pinckney, Nathaniel and Deng, Chenhui and Ho, Chia-Tung and Tsai, Yun-Da and Liu, Mingjie and Zhou, Wenfei and Khailany, Brucek and Ren, Haoxing},
  journal={arXiv preprint arXiv:2506.14074},
  year={2025}
}

@misc{openai_gpt5,
  title        = {GPT-5 Codex},
  author       = {{OpenAI}},
  howpublished = {OpenAI model documentation},
  note         = {Accessed: 2026-02-24}
}

@inproceedings{
liu2025deeprtl,
title={Deep{RTL}: Bridging Verilog Understanding and Generation with a Unified Representation Model},
author={Yi Liu and Changran XU and Yunhao Zhou and Zeju Li and Qiang Xu},
booktitle={The Thirteenth International Conference on Learning Representations},
year={2025},
url={https://openreview.net/forum?id=2hcfoCHKoB}
}

@inproceedings{liu2025deeprtl2,
  title={Deeprtl2: A versatile model for rtl-related tasks},
  author={Liu, Yi and Zhang, Hongji and Zhou, Yunhao and Shi, Zhengyuan and Xu, Changran and Xu, Qiang},
  booktitle={Findings of the Association for Computational Linguistics: ACL 2025},
  pages={6485--6500},
  year={2025}
}

@inproceedings{li2025specllm,
  title={Specllm: Exploring generation and review of vlsi design specification with large language model},
  author={Li, Mengming and Fang, Wenji and Zhang, Qijun and Xie, Zhiyao},
  booktitle={2025 International Symposium of Electronics Design Automation (ISEDA)},
  pages={749--755},
  year={2025},
  organization={IEEE}
}

@article{huang2025assessing,
  title={Assessing Large Language Models in Generating RTL Design Specifications},
  author={Huang, Hung-Ming and Yang, Yu-Hsin and Chang, Fu-Chieh and Hsu, Yun-Chia and Lin, Yin-Yu and Tsai, Ming-Fang and Yang, Chun-Chih and Wu, Pei-Yuan},
  journal={arXiv preprint arXiv:2512.00045},
  year={2025}
}

@inproceedings{
lee2026intrinsic,
title={Intrinsic Self-Correction in {LLM}s: Towards Explainable Prompting via Mechanistic Interpretability},
author={Yu-Ting Lee and Fu-Chieh Chang and Hui-Ying Shih and Pei-Yuan Wu},
booktitle={4th Deployable AI Workshop},
year={2026},
url={https://openreview.net/forum?id=74CcxRAnsp}
}

@inproceedings{
chang2025unveiling,
title={Unveiling the Latent Directions of Reflection in Large Language Models},
author={Fu-Chieh Chang and Yu-Ting Lee and Pei-Yuan Wu},
booktitle={Mechanistic Interpretability Workshop at NeurIPS 2025},
year={2025},
url={https://openreview.net/forum?id=e7YlRBe5Ra}
}

@inproceedings{
yubeaton2025verithoughts,
title={VeriThoughts: Enabling Automated Verilog Code Generation using Reasoning and Formal Verification},
author={Patrick Yubeaton and Andre Nakkab and Weihua Xiao and Luca Collini and Ramesh Karri and Chinmay Hegde and Siddharth Garg},
booktitle={The Thirty-ninth Annual Conference on Neural Information Processing Systems Datasets and Benchmarks Track},
year={2025},
url={https://openreview.net/forum?id=3Z8fWHKqlu}
}

@misc{jin2025realbenchbenchmarkingveriloggeneration,
      title={RealBench: Benchmarking Verilog Generation Models with Real-World IP Designs}, 
      author={Pengwei Jin and Di Huang and Chongxiao Li and Shuyao Cheng and Yang Zhao and Xinyao Zheng and Jiaguo Zhu and Shuyi Xing and Bohan Dou and Rui Zhang and Zidong Du and Qi Guo and Xing Hu},
      year={2025},
      eprint={2507.16200},
      archivePrefix={arXiv},
      primaryClass={cs.LG},
      url={https://arxiv.org/abs/2507.16200}, 
}

@article{shah2025rethinking,
  title={Rethinking reflection in pre-training},
  author={Shah, Darsh J and Rushton, Peter and Singla, Somanshu and Parmar, Mohit and Smith, Kurt and Vanjani, Yash and Vaswani, Ashish and Chaluvaraju, Adarsh and Hojel, Andrew and Ma, Andrew and others},
  journal={arXiv preprint arXiv:2504.04022},
  year={2025}
}

@inproceedings{wu2024autogen,
  title={Autogen: Enabling next-gen LLM applications via multi-agent conversations},
  author={Wu, Qingyun and Bansal, Gagan and Zhang, Jieyu and Wu, Yiran and Li, Beibin and Zhu, Erkang and Jiang, Li and Zhang, Xiaoyun and Zhang, Shaokun and Liu, Jiale and others},
  booktitle={First conference on language modeling},
  year={2024}
}

@inproceedings{zhang2024codeagent,
  title={Codeagent: Enhancing code generation with tool-integrated agent systems for real-world repo-level coding challenges},
  author={Zhang, Kechi and Li, Jia and Li, Ge and Shi, Xianjie and Jin, Zhi},
  booktitle={Proceedings of the 62nd Annual Meeting of the Association for Computational Linguistics (Volume 1: Long Papers)},
  pages={13643--13658},
  year={2024}
}

@inproceedings{wu-etal-2024-toolplanner,
    title = "{T}ool{P}lanner: A Tool Augmented {LLM} for Multi Granularity Instructions with Path Planning and Feedback",
    author = "Wu, Qinzhuo  and
      Liu, Wei  and
      Luan, Jian  and
      Wang, Bin",
    booktitle = "Proceedings of the 2024 Conference on Empirical Methods in Natural Language Processing",
    month = nov,
    year = "2024",
    publisher = "Association for Computational Linguistics",
    url = "https://aclanthology.org/2024.emnlp-main.1018/",
    doi = "10.18653/v1/2024.emnlp-main.1018",
    pages = "18315--18339",
}

@misc{
he2026refining,
title={Refining Specs For {LLM}-Based {RTL} Agile Design},
author={Sirui He and Chujie Chen and Xiang Zheng and Zhihang Liu and Cong Wang},
year={2026},
url={https://openreview.net/forum?id=1FADg2UNPn}
}

@inproceedings{yu2025spec2rtl,
  title={Spec2rtl-agent: Automated hardware code generation from complex specifications using llm agent systems},
  author={Yu, Zhongzhi and Liu, Mingjie and Zimmer, Michael and Celine, Yingyan and Liu, Yong and Ren, Haoxing},
  booktitle={2025 IEEE International Conference on LLM-Aided Design (ICLAD)},
  pages={37--43},
  year={2025},
  organization={IEEE}
}

@inproceedings{10.1109/DAC63849.2025.11133191,
author = {Zhao, Yujie and Zhang, Hejia and Huang, Hanxian and Yu, Zhongming and Zhao, Jishen},
title = {MAGE: A Multi-Agent Engine for Automated RTL Code Generation},
year = {2025},
isbn = {9798331503048},
publisher = {IEEE Press},
url = {https://doi.org/10.1109/DAC63849.2025.11133191},
doi = {10.1109/DAC63849.2025.11133191},
booktitle = {Proceedings of the 62nd Annual ACM/IEEE Design Automation Conference},
articleno = {361},
numpages = {7},
location = {San Francisco, California, United States},
series = {DAC '25}
}

\end{document}